\documentclass[12pt]{article}
\usepackage{amssymb}
\usepackage{epsfig}

\catcode`\@=11
\def\numberbysection{\@addtoreset{equation}{section}
        \def\theequation{\thesection.\arabic{equation}}}

\def\beq{\begin{equation}}
\def\eeq{\end{equation}}
\numberbysection
\begin{document}
\begin{titlepage}
\begin{center}
\hfill  \\
\vskip 1.in {\Large \bf Remarks on the harmonic oscillator with a minimal position uncertainty} \vskip 0.5in P. Valtancoli
\\[.2in]
{\em Dipartimento di Fisica, Polo Scientifico Universit\'a di Firenze \\
and INFN, Sezione di Firenze (Italy)\\
Via G. Sansone 1, 50019 Sesto Fiorentino, Italy}
\end{center}
\vskip .5in
\begin{abstract}
We show that this problem gives rise to the same differential equation of a well known
potential of ordinary quantum mechanics. However there is a subtle difference in the choice of the parameters of the hypergeometric function solving the differential equation which changes the physical discussion of the spectrum.
\end{abstract}
\medskip
\end{titlepage}
\pagenumbering{arabic}
\section{Introduction}

The non-renormalizability of quantum gravity requires the introduction of an intelligent $UV$ cutoff at the Planck scale, like non-commutative algebras. However this program is not easy to make explicit because many of them break basic principles of a quantum field theory like Lorentz invariance and unitarity.

It has been proposed that an effective cutoff in the ultraviolet should quantum theoretically be described as a non-zero minimal uncertainty $\Delta x_0$ in position measurements \cite{1}-\cite{2}. Technically it is necessary to require that there is no minimal uncertainty in momentum in order to use a continuous representation.

Recently it has also been found that this type of non-commutative cutoff is compatible with Lorentz invariance \cite{3}, in particular the Snyder geometry \cite{4}( at the price of loosing translational invariance ).

Since there are experimental observations pointing out that Lorentz invariance is preserved even at the Planck scale \cite{5}, this theoretical framework is worth being investigated.

In general it has been studied how this intelligent cutoff affects the underlying quantum mechanical structure. The most interesting example is the harmonic oscillator with a minimal position uncertainty, which has been solved in \cite{1}.

In this letter we point out that their solution leads to the same differential equation of a well known potential of quantum mechanics. We show that the difference between the two problems is in a different sign of the parameters of the hypergeometric functions solving the differential equation leading to a different physical spectrum.

\section{Minimal position uncertainty}

While in ordinary quantum mechanics $\Delta x$ can be made arbitrarily small, this is no longer the case if the following relation holds:

\beq \Delta x \Delta p \geq \frac{\hbar}{2} ( 1 + \beta ( \Delta p )^2 + \gamma ) \label{21}\eeq

The following Heisenberg algebra generated by $\hat{x}$ and $\hat{p}$ obeying the commutation relations

\beq [ \hat{x}, \hat{p} ] = i \hbar ( 1 + \beta \ \hat{p}^2 ) \label{22}\eeq

underlies the uncertainty relation \label{21} with $\gamma = \beta < p >^2$, from which

\beq \Delta p = \frac{\Delta x}{\hbar \beta } \pm \sqrt{ \left( \frac{\Delta x}{\hbar \beta} \right)^2 - \frac{1}{\beta} - < p >^2 } \label{23}\eeq

we can read off the minimal position uncertainty

\beq \Delta x_{min} ( <p> ) = \hbar \sqrt{\beta} \sqrt{ 1 + \beta < p >^2 } \geq  \hbar \sqrt{\beta} \label{24}\eeq

In the algebra (\ref{22}) there is no nonvanishing minimal uncertainty in momentum. The Heisenberg algebra (\ref{22}) can be represented continuously on momentum space wave functions $\psi(p) = < p | \psi >$ as

\begin{eqnarray}
\ \hat{p} \psi(p) \ & = & \ p \psi(p) \nonumber \\
\ \hat{x} \psi(p) \ & = & \ i \hbar ( 1 + \beta p^2 ) \frac{\partial}{\partial p} \psi(p)
\label{25}
\end{eqnarray}

The position and momentum operators are symmetric on the domain $S_{\infty}$ with respect to the following modified scalar product:

\beq < \psi | \phi > \ = \ \int^{+\infty}_{-\infty} \ \frac{ dp }{1 \ + \ \beta p^2} \ \psi^*(p) \phi (p) \label{26} \eeq

The identity operator can thus be expanded as

\beq 1 = \int^{+\infty}_{-\infty} \ \frac{ dp }{1 \ + \ \beta p^2} \ | p > < p | \label{27}\eeq

and the scalar product of momentum eigenstates is:

\beq < p | p' > \ = \ ( 1 + \beta p^2 ) \ \delta ( p - p' ) \label{28}\eeq

The position operator is no longer essentially self-adjoint but has a one parameter family of self-adjoint extensions. This means that even if formal eigenvectors for the position operator exist, they are not physical states since they have infinite energy.

\section{The harmonic oscillator}

From the expression for the Hamiltonian

\beq H \ = \ \frac{\hat{p}^2}{2m} \ + \ m \omega^2 \frac{\hat{x}^2}{2} \label{31}\eeq

and the representation for $\hat{x}$ and $\hat{p}$ in the $p$-space (\ref{25}) we get the following form for the stationary state Schrodinger equation:

\beq \frac{d^2 \psi(p)}{d p^2}  + \frac{2\beta p}{1+\beta p^2} \
\frac{d \psi(p)}{d p} + \frac{ 4 \beta ( q - \beta r p^2 )}{(1+\beta p^2)^2} \ \psi(p) = 0 \label{32}
\eeq

where

 \beq q = \frac{E}{2m \hbar^2 \omega^2 \beta }  \ \ \ \  r = \frac{1}{4 \beta^2 m^2 \hbar^2 \omega^2 } \label{33}\eeq

  and $E$ is the energy.

In order to find the explicit solution it is useful to introduce a new variable $z$ in terms of which the poles coincide with those of the hypergeometric function, i.e. $ 0, 1, \infty$ :

\beq \frac{d^2 \psi(z)}{d z^2}  + \frac{ 2z-1 }{ z(z-1) } \
\frac{d \psi(z)}{d z} - \frac{( q + r ( 1 - 2 z )^2 )}{ z^2 ( z-1 )^2} \ \psi(z) = 0
\ \ \ \ z = \frac{1}{2} + i \frac{\sqrt{\beta}}{2} p \label{34}\eeq

This type of differential equation is not new in physics since it is related to a well known potential of ordinary quantum mechanics:

\beq V(x) = - \frac{U_0}{ch^2 \alpha x}  \label{35}\eeq

The corresponding eigenvalue problem is, in the $x$-space representation, given by:

\beq \frac{d^2 \psi(x) }{dx^2} + \frac{2m}{\hbar^2} \left( E + \frac{U_0 }{ch^2 \alpha x} \right) \psi(x) = 0 \label{36}\eeq

The analysis of the spectrum is rather clear in this case. The positive energy spectrum is continuous and the negative energy spectrum is discrete. Let us remember the discussion in the last case. By changing variables $ \xi = th ( \alpha x ) $ and introducing the notations:

\beq \epsilon = \frac{\sqrt{-2mE}}{\hbar \alpha} \ \ \ \ \ s(s+1) \ = \ \frac{2m U_0}{\hbar^2 \alpha^2} \label{37} \eeq

we obtain the differential equation:

\beq \frac{d}{d\xi} \left[ ( 1-\xi^2 ) \frac{d\psi}{d\xi} \right] + \left[ s(s+1) - \frac{\epsilon^2}{1-\xi^2} \right] \psi = 0 \label{38}\eeq

To reach the hypergeometric function it is necessary the substitution:

\begin{eqnarray}
\psi & = & (1-\xi^2)^{\epsilon / 2} w( \xi ) \nonumber \\
u & = & \frac{1}{2} ( 1 - \xi ) \label{39}
\end{eqnarray}

We end up with the hypergeometric equation:

\beq u(1-u) w'' + ( \epsilon + 1 ) ( 1 - 2u ) w' - ( \epsilon - s ) ( \epsilon + s + 1 ) w = 0 \label{310}\eeq

The finite solution for $\xi = 1 ( x = +\infty )$ is

\beq \psi = ( 1 - \xi^2 )^{\epsilon / 2} \ {}_2 F_1 \left[ \epsilon- s, \epsilon+s+1, \epsilon+1 , \frac{1-\xi}{2} \right] \label{311}\eeq

The condition of finiteness of $\psi$ at $\xi=-1 ( x= - \infty )$ requires $\epsilon-s = - n$ ( where n is an integer ), then the hypergeometric function reduces to a polynomial of degree $n$.

In this way, the energy levels are determined by the condition

\beq s-\epsilon = n \label{312}\eeq

from which

\beq E_n = - \frac{\hbar^2 \alpha^2}{8m} \left[ - ( 1 + 2n ) + \sqrt{1 + \frac{8m U_0}{\alpha^2 \hbar^2} } \right]^2 \label{313}\eeq

The number of discrete energy levels is finite and determined by the condition $\epsilon > 0$ i.e.

\beq n < s \label{314}\eeq

The differential equation of the quantum mechanical problem (\ref{38}) compared with the noncommutative problem (\ref{34}) implies the following relation between the parameters

\beq q = \frac{\epsilon^2- s(s+1)}{4} \ \ \ \ r = \frac{s(s+1)}{4} \ \ \ \ \epsilon = 2 \sqrt{ q+r } \label{315}\eeq

In the differential equation it appears only the combination $\epsilon^2$. We therefore want to clarify that while in the quantum mechanical problem the natural choice is the solution with the positive sign $+\epsilon$, in the non-commutative problem the right choice turns out to be $-\epsilon$. In fact the non-commutative problem requires the substitution

\beq \psi = (1-\xi^2)^{-\epsilon / 2} w(\xi) \ \ \ \ u = \frac{1}{2} ( 1-\xi ) \label{316}\eeq

which implies the following solution in the $p$ variables ( $\xi = - i \sqrt{\beta} p $ ):

\beq \psi (p) = \frac{1}{
(1+\beta p^2)^{\sqrt{q+r}}} {}_2 F_1 \left( -\epsilon -s, -\epsilon+s+1, 1- \epsilon , \frac{1}{2} + i\frac{\sqrt{\beta}}{2} p \right) \label{317}\eeq

The quantization condition is now

\beq \epsilon + s = n \label{318}\eeq

and for $q$ we obtain

\beq q = \frac{(n-s)^2}{4} - r \label{319}\eeq

The requirement $\epsilon > 0$ i.e. $ n-s > 0 $ is always satisfied for every integer $n$
by choosing also for $s$ the negative solution

\beq s = -\frac{1}{2} - \frac{\sqrt{1+16 r}}{2} \label{320}\eeq

The number of energy levels is now infinite and not subjected to any condition

\begin{eqnarray}
E_{level} & \sim & \hbar \omega \frac{q}{\sqrt{r}} = \hbar \omega \left( \frac{(n-s)^2}{4\sqrt{r}} - \sqrt{r} \right) = \nonumber \\
& = & \hbar \omega \left[ \left(n+\frac{1}{2}\right) \left( \sqrt{ 1 + \frac{1}{16r} } + \frac{1}{4\sqrt{r}} \right) + \frac{n^2}{4\sqrt{r}} \right]
\label{321}\end{eqnarray}

For $\beta \rightarrow 0 \ ( r\rightarrow \infty )$ we recover the harmonic oscillator energy levels.

The conclusion is that it is enough to choose a different sign ( $\pm \epsilon $ ) in the solution of the same differential equation (\ref{34}) to reach completely different physical spectra.


\begin{thebibliography}{999}
\bibitem{1} A. Kempf, G. Mangano, R. B. Mann,  Phys. Rev. {\bf D 52} (1995) 1108, hep-th/9412167.
\bibitem{2} A. Kempf, G. Mangano, Phys. Rev {\bf D 55} (1997) 7909, hep-th/9612084.
\bibitem{3} A. Bose, ArXiv:1011.5234v1[physics.gen-ph].
\bibitem{4} H. S. Snyder, Phys. Rev {\bf 71} (1947) 38.
\bibitem{5} V. A. Kostelecky, N. Russell, Rev. Mod. Phys. {\bf 83} (2011) 11, ArXiv:0801.0287v3[hep-ph].
\end{thebibliography}
\end{document}